\documentclass[epj]{svjour}
\usepackage{subeqn,graphicx,subfigure}
\begin{document}
\title{Centrality-dependence of Particle Production at RHIC and
the Combinational Approach}

\author{Bhaskar De\thanks{e-mail: bhaskar$\_$r@isical.ac.in}  and S.
Bhattacharyya\thanks{e-mail: bsubrata@isical.ac.in (Communicating
Author).}}

\institute{Physics
and Applied Mathematics Unit(PAMU),\\
Indian Statistical Institute, Kolkata - 700108, India.}

\date{Received: to be given by the Editor}

\abstract{
The newly proposed combinational approach, called the grand
combinational model (GCM), as would be described in detail in the
text, is still under our careful scrutiny. By applying it, we have
attempted to analyze here the characteristics of both the
transverse momentum($p_{\rm{T}}$)-, and centrality-dependence of
production of the main varieties of the secondaries measured in
$AuAu$ collisions at BNL-RHIC at both $\sqrt{s_{\rm{NN}}}=130$ GeV and
$\sqrt{s_{\rm{NN}}}=200$ GeV by PHENIX Collaboration. Besides, with the
help of it, we have also investigated the nature of
centrality-dependence of the average transverse momenta of the
various major categories of particles in $AuAu$ collisions at
RHIC. The model seems to survive quite smoothly the acid tests of
the latest PHENIX data, as it accommodates data modestly well on
these twin aspects. The study reveals a kind of universality of
nature of the hadronic secondaries and also of the basic particle
and nuclear interactions at high energies. However, in the end, we
precisely point out both the strengths and limitations of the specific
model under consideration here.}
\authorrunning{B. De et al }
\titlerunning{Centrality-dependence of Particle Production at RHIC ...}
\maketitle

\newpage

\section{Introduction}
In the recent past the PHENIX Collaboration conducted exhaustive
measurements of the particle-yields in $AuAu$ collisions at
RHIC-BNL at very high ranges of transverse momentum values and
also at various centralities of the collisions. The data-sets on
both identified centrality- dependent charged hadron spectra vs.
transverse momenta and also on the average transverse
momentum-values $(<p_{\rm{T}}>)$ of the various secondaries vs.
the centralities of the collisions\cite{Chujo1,Mioduszewski1} were
indicated by the graphical plots shown in various figures
presented here. These measurements were mainly aimed at checking
some predictions from the Standard Model(SM) on (i) the
suppression of particle yields at large transverse momenta and
(ii) the dependence of the degree of suppression on the centrality
of the collision. In the framework of the standard model, all the
secondary particles are produced by jets. These jets are likely to
suffer significant energy loss via `gluon' radiation, while
passing through the hot dense medium arising out of the
anticipated QGP formation. So, according to this view, the
particle yield at large $p_{\rm{T}}$ should be suppressed to a
considerable degree. Secondly, since the amount of energy loss is
a function of the density and the path length through such a
projected QGP media, the suppression effect is predicted to be
dependent also on the centrality of the collision\cite{Jia1}. All
the theoretical postulates and predictions serve as the background
for such intense attempts at measurements.
\par
But in our approach we would maintain a degree of neutrality to
such standard theoretical views and would concern ourselves mainly
with the results of measurements with the ulterior motive of
checking and testing an alternative approach, called here the
Combinational Approach or the grand combination of models(GCM)
which would be utilized here to explain the sets of data on: (i)
identified charged hadron spectra vs. $p_{\rm{T}}$-values and (ii) the
values of average transverse momentum, denoted by $<p_{\rm{T}}>$, vs. the
different centrality-ranges of the $AuAu$ reaction in particular.
\par
The work is organized as follows: In Section 2 we offer an outline
of the GCM to be made use of in the present study. The Section 3
contains and depicts the concrete results obtained on the basis of
this model in the tabular forms and the graphical plots. The last
section presents, as usual, the summary and final comments.

\section{The Model: A Sketch}
Following the suggestion of Faessler\cite{Faessler1} and the work
of Peitzmann\cite{Peitzmann1} and also of Schmidt and
Schukraft\cite{Schmidt1}, we propose here a generalized empirical
relationship between the inclusive cross-sections for any variety
of the secondaries ($Q$), such as hadrons, pions, kaons or
proton/antiprotpons, produced in nucleon(N)-nucleon(N) collision
and that for nucleus (A)-nucleus(B) collision as given below:

\begin{equation}
\displaystyle{E\frac{d^3  \sigma}{dp^3} ~ (AB \rightarrow Q X) ~
\sim ~ (AB)^{\phi(y, ~ p_{\rm{T}})} ~ E\frac{d^3  \sigma}{dp^3} ~ (PP
\rightarrow Q X) ~,} \label{eqn1}
\end{equation}

where $\phi(y, ~ p_{\rm{T}})$ could be expressed in the factorization
form, $\phi(y, ~ p_{\rm{T}}) = f(y) ~ g(p_{\rm{T}})$; and the product, $AB$ on
the right hand side of the above equation is the product of mass
numbers of the two nuclei participating in the collisions at high
energies, of which one will be the projectile and the other one
the target.
\par
While investigating a specific nature of dependence of
the two variables($y$ and $p_{\rm{T}}$), either of them is assumed to
remain averaged or with definite values. Speaking in clearer
terms, if and when $p_{\rm{T}}$-dependence is studied by experimental
group, the rapidity factor is integrated over certain limits
and is absorbed in the normalization factor. So, the formula turns into

\begin{equation}
\displaystyle{E\frac{d^3 \sigma}{dp^3} ~ (AB \rightarrow Q X) ~
\sim ~ (AB)^{g(p_{\rm{T}})} ~ E\frac{d^3  \sigma}{dp^3} ~ (PP \rightarrow
Q X) ~,} \label{eqn2}
\end{equation}

The main bulk of work, thus, converges to the making of an
appropriate choice of form for $g(p_{\rm{T}})$. And the necessary choices
are to be made on the basis of certain premises and physical
considerations which do not violate the canons of high energy
particle interactions.
\par
The expression for inclusive cross-section of $Q$ in proton-proton
scattering at high energies occurring in Eqn.(2) could be chosen
in the form suggested first by Hagedorn\cite{Hagedorn1}:
\begin{equation}
\displaystyle{ E\frac{d^3  \sigma}{dp^3} ~ (PP \rightarrow Q X) ~
= ~ C_1 ~ ( ~ 1 ~ + ~ \frac{p_{\rm{T}}}{p_0})^{-n} ~ ,} \label{eqn3}
\end{equation}

where $C_1$ is the normalization constant, and $p_o$, $n$ are
interaction-dependent chosen phenomenological parameters for which
the values are to be obtained by the method of fitting the spectra
in $PP$ interaction.
\par
The final working formula for the nucleus-nucleus collisions is
now being proposed here in the form given below:

\begin{equation}
\begin{array}{lcl}
E\frac{d^3  \sigma}{dp^3}  (AB \rightarrow Q X)  & \propto & ~
(AB)^{(\epsilon  +  \alpha  p_{\rm{T}}   -  \beta p_{\rm{T}}^2)} ~
E\frac{d^3 \sigma}{dp^3}  (PP \rightarrow Q X) \\
& \propto & ~ (AB)^{(\epsilon ~ + ~ \alpha  p_{\rm{T}}  ~ - ~ \beta
p_{\rm{T}}^2)} ~ (1 ~ + ~ \frac{p_{\rm{T}}}{p_0})^{-n}  ,  \label{eqn4}
\end{array}
\end{equation}

with $ g(p_{\rm{T}}) ~ = ~ (\epsilon ~ + ~ \alpha  p_{\rm{T}}  ~ - ~ \beta
p_{\rm{T}}^2)$, where this suggestion of quadratic parametrization for
$g(p_{\rm{T}})$ is exclusively made by us and is called hereafter
De-Bhattacharyya parametrization(DBP). In the above expression
$\epsilon$, $\alpha$ and $\beta$ are constants for a specific pair
of projectile and target.
\par
Earlier experimental works\cite{Aggarwal1,Albrecht1,Antreasyan1}
showed that $g(p_{\rm{T}})$ is less than unity in the
$p_{\rm{T}}$-domain, $p_{\rm{T}}<1.5$ GeV/c. Besides, it was also
observed that the parameter $\epsilon$, which gives the value of
$g(p_{\rm{T}})$ at $p_{\rm{T}}=0$, is also less than one and this
value differs from collision to collision. The other two
parameters $\alpha$ and $\beta$ essentially determine the nature
of curvature of $g(p_{\rm{T}})$. However, in the present context,
precise determination of $\epsilon$ is not possible for the
following understated reasons:
\par
(i) To make our point let us recast the expression for (4) in the
form given below:
\begin{equation}
\displaystyle{E\frac{d^3\sigma}{dp^3}(AB \rightarrow Q X)
\approx  C_2  (AB)^\epsilon  (AB)^{(\alpha p_{\rm{T}} - \beta p_{\rm{T}}^2)}
 (  1 + \frac{p_{\rm{T}}}{p_0}  )^{-n}}\label{eqn5}
\end{equation}
where $C_2$ is the normalization term which has a dependence
either on the rapidity or on the rapidity density of the $Q$ and
which also absorbs the previous constant term,$C_1$ as well.
\par
Quite obviously, we have adopted here the method of fitting. Now,
in Eqn.(5) one finds that there are two constant terms $C_2$ and
$\epsilon$ which are neither the coefficients nor the exponent
terms of any function of the variable, $p_{\rm{T}}$. And as $\epsilon$ is
a constant for a specific collision at a specific energy, the
product of the two terms $C_2$ and $(AB)^\epsilon$ appears as just
a new constant. And, it will just not be possible to obtain
fit-values simultaneously for two constants of the above types by
the method of fitting.
\par
(ii) From Eqn.(2) the nature of $g(p_{\rm{T}})$ can easily be determined
by calculating the ratio of the logarithm of the ratios of
nuclear-to-$PP$ collision and the logarithm of the product $AB$.
Thus, one can measure $\epsilon$ from the intercept of $g(p_{\rm{T}})$
along y-axis as soon as one gets the values of
$E\frac{d^3\sigma}{dp^3}$ for any specific secondary production in
both $AB$ collision and $PP$ collision at the same c.m. energy. In
the present study we have tried to consider the $AuAu$ collision
system in various centrality bins at two different c.m.energis. In
order to do so, we have to consider the data on normalized
versions of $E\frac{d^3\sigma}{dp^3}$ for any secondary particle
produced in this collision system for which no clear
$E\frac{d^3\sigma}{dp^3}$-data is available to us. Furthermore,
from these normalized versions we can/could not extract the
appropriate values of $E\frac{d^3\sigma}{dp^3}$ as the
normalization terms, total inclusive cross-sections$(\sigma_{in})$
etc., for this collision system at all centrality-bins cannot
always be readily obtained. Besides, it will also not be possible
to get readily the data on inclusive spectra for $PP$ collisions
at all c.m.energies.
\par
In order to sidetrack these difficulties and also to build up an
escape-route, we have concentrated here almost wholly to the
values of $\alpha$ and $\beta$ for various collision systems and
the resultant effects of $C_2$ and $\epsilon$ have been absorbed
into a single constant term $C_3$. Hence, the final expression
becomes
\begin{equation}
\displaystyle{E\frac{d^3\sigma}{dp^3}(AB \rightarrow Q X) ~
\approx ~ C_3 ~ (AB)^{(\alpha p_{\rm{T}} - \beta p_{\rm{T}}^2)} ~ ( ~ 1 ~ +
\frac{p_{\rm{T}}}{p_0} ~ )^{-n}}\label{eqn6}
\end{equation}
with $C_3 = C_2 (AB)^\epsilon$.
\par
The exponent factor term $\alpha p_{\rm{T}} - \beta p_{\rm{T}}^2$
obviously represents here $[g(p_{\rm{T}})-\epsilon]$ instead of
$g(p_{\rm{T}})$ alone. The expression(6) given above is the
physical embodiment of what we have termed to be the grand
combination of models(GCM) that has been utilized here. The
results of $PP$ scattering are obtained in the above on the basis
of eqn.(3) provided by Hagedorn's model(HM);  and the route for
converting the results of $NN$ to $NA$ or $AB$ collisions is built
up by the Peitzmann's approach(PA) represented by expression(2).
The further input is the De-Bhattacharyya parametrization for the
nature of the exponent. Thus, the GCM is the combination of HM, PA
and the DBP, all of which are used here.
\par
And the choice of this form of parametrization for the power of
the exponent in eqn.(4) is not altogether a coincidence. In
dealing with the EMC effect in the lepton-nucleus collisions, one
of the authors here(SB),\cite{SB1} made use of a polynomial form
of $A$-dependence with a variable $x$ which is a variant of $x_F$
(the Feynman Scaling variable). This gives us a clue to make a
similar choice for both $g(p_{\rm{T}})$ and $f(y)$ variable(s) in
each case separately. In the recent times, De-Bhattacharyya
parametrization is being extensively applied to interpret the
measured data on the various aspects\cite{De1,De2,De3,De4} of the
particle-nucleus and nucleus-nucleus interactions at high
energies. In the recent past Hwa et. al.\cite{Hwa1} also made use
of this sort of relationship in a somewhat different context. The
underlying physics implications of this parametrization stem
mainly from the expression(4) which could be identified as a clear
mechanism for switch-over of the results obtained for
nucleon-nucleon($PP$) collision to those for nucleus-nucleus
interactions at high energies in a direct and straightforward
manner. The polynomial exponent of the product term on $AB$ takes
care of the totality of the nuclear effects.
\par
For the sake of clarity and confirmation, let us further emphasize
a point here very categorically. It is to be noted that this
model(GCM) containing all the Eqns.(4), (5) and (6) was described
in some detail earlier and was made use of in analyzing extensive
sets of data in the previous publications\cite{De1,De2,De4} by the
same authors. And in verifying the validity of this model further,
the purpose here is to apply the same model to some other
problematical aspects of data which we would dwell upon in the
subsequent sections. Before taking them up, let us state a point.
In some previous works\cite{De2,De3,De4} we tried to provide some
sort of physical interpretations for some of the parameters used
in the present work. But, those explanations were only of
suggestive nature. Besides, obviously, they are not complete and
sufficient, for which we have chosen not to reiterate them here
once more.
\section{Presentation of Results}
Obviously the GCM is the model of our choice here. The inclusive
spectra for production of the main varieties of various
secondaries produced in $AuAu$ collisions at RHIC at both
$\sqrt{s_{\rm{NN}}}=130$ GeV and $\sqrt{s_{\rm{NN}}}=200$ GeV and also at
various centrality values have been worked out here
phenomenologically and shown in the several diagrams. The values
of $p_0$ and $n$, occurring in eqn.(3), which are essentially the
contribution of $PP$ collisions to the nucleus-nucleus collisions
for the same secondaries produced at the same c.m.energies per
nucleon, have been introduced by the following relationships:

\begin{equation}
\displaystyle{ p_0(\sqrt{s}) ~ = ~ a ~ + ~ \frac{b}{\sqrt{\frac{s_{\rm{NN}}}{{\rm{GeV}}^2}}
~ \ln({\sqrt{\frac{s_{\rm{NN}}}{{\rm{GeV}}^2}})}}} \label{eqn7}
\end{equation}

\begin{equation}
\displaystyle{ n(\sqrt{s}) ~ =  ~ \acute{a} ~ + ~
\frac{\acute{b}}{\ln^2({\sqrt{\frac{s_{\rm{NN}}}{{\rm{GeV}}^2}}})}} \label{eqn8}
\end{equation}

These are products of just empirical analyzes made earlier and
reported in some of our previous works\cite{De1,De2}. The actually
used values of the arbitrary parameters, $a,b,\acute{a}$ and
$\acute{b}$ for various secondary particles are given in Table-1.
\par
The obtained values of the average yields of hadrons at various
centralities have been depicted in Fig.1. The left panel is for
$\sqrt{s_{\rm{NN}}}=130$ GeV and the right panel is for
$\sqrt{s_{\rm{NN}}}=200$ GeV. The Fig.2 describes pion production on a
charge-neutral and average basis at both the energies. Similar is
case with kaons in Fig.3. The diagrams in Fig.4 reproduce data on
proton production in $AuAu$ collisions at the two RHIC energies.
The solid curves in Fig.5 display the theoretical yields of
antiprotons in $AuAu$ reaction at various values of the centrality
of the collision against measured data. The values of the
parameters $\alpha$ and $\beta$ have been initially chosen with
the singular motivation of obtaining satisfactory fits to the
data, though finally even this arbitrariness has led to some
revelation of the specific nature of $\alpha$ and $\beta$ as are
shown in the plots of Fig.6 for the various secondaries in several
panels.
\par
The values of $\alpha$ and $\beta$ to be used in obtaining our
model-based results are shown in the different tables(Table 2 -
Table 11). The extreme left columns in all of them contain
information about the centrality of the reaction and the extreme
right ones offer the $\chi^2$/ndf values. Even for the cases of
proton and antiproton production, wherein the data suffer a high
degree of uncertainty, the $\chi^2$/ndf values are modestly
satisfactory. The systematic trends of the used values of $\alpha$
and $\beta$ depict a harmony of their nature which have been
hinted by Table 12 and Table 13 and represented by the sets of
diagrams in Fig.6. The solid lines in Fig.6 provide the
phenomenological fits which can be expressed by a common
relationship of the form given below,

\begin{equation}
\displaystyle{\alpha(N_{\rm{part}}),\beta(N_{\rm{part}}) ~ = ~ R ~ + ~ S ~
\ln{(N_{\rm{part}})}}\label{eqn9}
\end{equation}

The different values of $R$ and $S$ for various secondaries are
given in Table 12 and Table 13.
\par
A comment is in order here in a preemptive manner. The values of
$\alpha$ and $\beta$ shown in our previous work \cite{De1} even on
$AuAu$ collision could be and are little different from what are
depicted here for the two reasons: (i) The $p_{\rm{T}}$-range of
the detected secondaries in the previous work was limited mostly
in the region from 0.8 GeV/c to 3 GeV/c, whereas in the present
case both the full low $p_{\rm{T}}$ and a larger domain of high
$p_{\rm{T}}$ range for the secondaries(mainly charged hadrons) has
been covered. (ii) Secondly, in the former study\cite{De1} the
minimum bias event was studied in the main. On the contrary, the
present study has very much been centrality-specific and the data
for the various centrality-values of the $AuAu$ collision have
been served within a phenomenological framework.
\par
The average transverse momenta values for the different categories
of particles have, however, been worked out on the basis of the
following expression:

\begin{equation}
\displaystyle{<p_{\rm{T}}> ~ = ~ \frac{\int_{0}^{\infty} {p_{\rm{T}} ~
\frac{dN}{p_{\rm{T}} ~ dp_{\rm{T}}} ~ dp_{\rm{T}}^2}}{\int_{0}^{\infty}{\frac{dN}{p_{\rm{T}} ~
dp_{\rm{T}}} ~ dp_{\rm{T}}^2}}}\label{eqn10}
\end{equation}

The values of $\alpha$ and $\beta$ to be introduced for
$<p_{\rm{T}}>$-values are used in both energy-specific and
particle-specific manner with the help of eqn.(9). The GCM-based
results on $<p_{\rm{T}}>$ values are plotted in Fig.7. The different
centrality values, the particle-species and the interaction
energy-values are separately mentioned in each of the diagrams.

\section{Concluding Remarks}
The chosen model appears to present essentially a universal
approach in the sense that (i) it provides a unified description
of data on particle production in nuclear collisions in terms of
the basic $PP$ interaction; (ii) the method could be applied in an
integrated and uniform way without introduction of any artificial
divide between the so-called `soft'(low-$p_{\rm{T}}$) and
`hard'(large-$p_{\rm{T}}$) interactions; (iii) the general
approach remains valid, irrespective of whether the collisions are
central or peripheral; (iv) it has no model- or mechanism-specific
physical picture as the input and as the constraint as well; (v)
the values of $\alpha$ and $\beta$ which are the only arbitrary
parameters need to be assumed and they demand tuning and
adjustment on a case-to-case basis in an interaction-specific,
secondary-specific and centrality-specific manner. So, the model
seems to provide a universal, useful and economical description of
a large body of data on the high energy heavy ion collisions. The
model is useful, as it is seen to give a fair account of the vast
amount of data; and it is economical, because there are only two
arbitrary parameters alongwith one normalization term for the
general studies on heavy ion reactions at high energies.
\par
The agreements between the measured and/or extracted data and the
phenomenological outputs are quite satisfactory on an overall
basis of the $p_{\rm{T}}$-spectra and the $<p_{\rm{T}}>$ vs.
centrality diagrams. The only exception is the case of the average
transverse momenta values of kaons in $AuAu$ reaction at
$\sqrt{s_{\rm{NN}}}=200$ GeV. Though we cannot readily ascribe any
reason for such departure, there is a general observation that the
measurements related to any variety of the strange particles
suffer, in general, a higher degree of uncertainty. Besides, we
also fail to explain here how and why this phenomenological
approach works functionally so well. No clue to any concrete
physical reason arising out of some underlying dynamics of
particle and nuclear interactions could be immediately
ascertained. The harmony revealed in the natures of $\alpha$ and
$\beta$ versus various $N_{\rm{part}}$-values reflecting various
centrality of the reactions is certainly an interesting
observation from the present approach. The closeness of the values
of $S$ in the Table-12 and Table-13 at two different neighbourly
energies appears to indicate the fact that, beyond a certain value
of $N_{\rm{part}}$, the enhancement of the centrality of the
collision with the increase in the number of wounded nucleons,
i.e. $N_{\rm{part}}$ does not necessarily and appreciably raise
the values of $<p_{\rm{T}}>$s for any variety of the secondaries.
This brings out strong hints to what is called parton saturation
at and after a definite value of $N_{\rm{part}}$. In fine, in so
far as actual performance is concerned, the model has a modest
degree of success. However, one major drawback in applying this
approach is it's over-reliance on the availability of the measured
and dependable data-sets on the specific variety of the secondary
in $PP$ interaction at some definite energies and at certain
reasonable intervals in the energy-values in order to construct
the energy-dependence profile for some parameters to be used in
the model. Secondly, the final working formula for studying the
properties of nuclear collisions in the present work does neither
contain directly, nor exhibit any of the technicalities of the
nuclear geometry, e.g., the impact parameter(denoted generally by
$b$) or of the space-time evolution scenarios of the nuclear
collisions. The entirety of the nuclear effects is taken care of
by the simple product term $(AB)^{f(y,p_{\rm{T}})}$. This
simplicity of form could very well be viewed in a positive way in
favour of the model.

\bigskip

\begin{acknowledgement}
The authors would like to express their thankful gratitude to the
anonymous referee for his/her critical remarks and valuable
suggestions for improvement of an earlier draft of the manuscript.
\end{acknowledgement}


\newpage
\begin{table*}
\caption{Values of $a$,$b$,$a'$ and $b'$ used in eqn.(7) and in
eqn.(8) to obtain $p_0$ and $n$ for various secondaries.}
\centering
\begin{tabular}{lllll}
\hline Secondary Type & $a$ & $b$ & $\acute{a}$ & $\acute{b}$\\
\hline $h$,$\pi$ & 1.5 & 79.4 & 6.5 & 127\\
       $K$       & 1.6 & 103  & 3.6 & 161\\
       $P$       & 7   & 602  & 5   & 644\\
       $\bar{P}$  & 7   & 478  & 13  & 527\\
\hline
\end{tabular}
\end{table*}
\begin{table*}
\caption{Parameter values for charged hadrons(averaged) production
in $AuAu$ collision at $\sqrt{s_{\rm{NN}}}=$130 GeV} \centering
\begin{tabular}{lllll}
\hline Centrality & $C_3$ & $\alpha$(c/GeV) & $\beta$(c/GeV)$^2$ &
$\chi^2$/ndf\\
\hline $0-15\%$ & $1708\pm42$ & $0.10\pm0.01$ & $0.019\pm 0.002$ &
1.127\\
$5-10\%$ & $1487\pm39$ & $0.09\pm0.01$ & $0.017\pm 0.002$ &
1.281\\
$10-20\%$ & $1161\pm31$ & $0.088\pm0.004$ & $0.016\pm 0.001$ &
1.401\\
$20-30\%$ & $831\pm21$ & $0.077\pm0.003$ & $0.012\pm 0.001$ &
1.425\\
$30-40\%$ & $561\pm15$ & $0.073\pm0.003$ & $0.012\pm 0.002$ &
1.590\\
$40-60\%$ & $262\pm5$ & $0.070\pm0.002$ & $0.011\pm 0.001$ &
0.714\\
$60-80\%$ & $69\pm2$ & $0.063\pm0.003$ & $0.009\pm 0.001$ &
0.214\\
\hline
\end{tabular}
\end{table*}
\begin{table*}
\caption{Parameter values for production of charged
hadrons(averaged) in $AuAu$ collision at $\sqrt{s_{\rm{NN}}}=$200 GeV}
\centering
\begin{tabular}{lllll}
\hline Centrality & $C_3$ & $\alpha$(c/GeV) & $\beta$(c/GeV)$^2$ &
$\chi^2$/ndf\\
\hline $0-5\%$ & $3015\pm189$ & $0.072\pm0.006$ & $0.012\pm 0.002$
& 1.281\\
$10-15\%$ & $1962\pm120$ & $0.0741\pm0.006$ & $0.012\pm 0.002$ &
1.191\\
$20-30\%$ & $1230\pm111$ & $0.064\pm0.009$ & $0.009\pm 0.002$ &
1.847\\
$40-50\%$ & $547\pm32$ & $0.052\pm0.004$ & $0.006\pm 0.001$ &
0.943\\
$60-70\%$ & $171\pm8$ & $0.041\pm0.004$ & $0.005\pm 0.001$ &
0.403\\
$80-91\%$ & $262\pm5$ & $0.070\pm0.002$ & $0.011\pm 0.001$ &
0.521\\
\hline
\end{tabular}
\end{table*}
\begin{table*}
\caption{Parameter values for charged pions(averaged) produced in
$AuAu$ collision at $\sqrt{s_{\rm{NN}}}=$130 GeV} \centering
\begin{tabular}{lllll}
\hline Centrality & $C_3$ & $\alpha$(c/GeV) & $\beta$(c/GeV)$^2$ &
$\chi^2$/ndf\\
\hline $0-5\%$ & $1243\pm113$ & $0.07\pm0.02$ & $0.022\pm 0.010$
& 0.948\\
$5-15\%$ & $922\pm44$ & $0.07\pm0.01$ & $0.017\pm 0.005$ &
0.399\\
$15-30\%$ & $602\pm47$ & $0.066\pm0.016$ & $0.017\pm 0.008$ &
1.077\\
$30-60\%$ & $230\pm24$ & $0.062\pm0.022$ & $0.015\pm 0.010$ &
1.774\\
$60-90\%$ & $33\pm10$ & $0.046\pm0.010$ & $0.016\pm 0.006$ &
1.859\\
\hline
\end{tabular}
\end{table*}
\begin{table*}
\caption{Parameter values for charged pions(averaged) produced in
$AuAu$ collision at $\sqrt{s_{\rm{NN}}}=200$ GeV} \centering
\begin{tabular}{lllll}
\hline Centrality & $C_3$ & $\alpha$(c/GeV) & $\beta$(c/GeV)$^2$ &
$\chi^2$/ndf\\
\hline $0-5\%$ & $1375\pm44$ & $0.068\pm0.007$ & $0.024\pm 0.003$
& 0.896\\
$10-15\%$ & $1170\pm26$ & $0.048\pm0.005$ & $0.015\pm 0.002$ &
0.850\\
$20-30\%$ & $702\pm24$ & $0.055\pm0.007$ & $0.017\pm 0.003$ &
0.566\\
$30-40\%$ & $502\pm15$ & $0.039\pm0.006$ & $0.010\pm 0.003$ &
0.817\\
$40-50\%$ & $309\pm14$ & $0.038\pm0.008$ & $0.012\pm 0.003$ &
0.805\\
$50-60\%$ & $181\pm4$ & $0.030\pm0.004$ & $0.010\pm 0.002$ &
1.159\\
$60-70\%$ & $96\pm2$ & $0.022\pm0.004$ & $0.010\pm 0.003$ &
0.881\\
$70-80\%$ & $44\pm1$ & $0.022\pm0.006$ & $0.018\pm 0.005$ &
1.623\\
$80-91\%$ & $23\pm1$ & $0.017\pm0.004$ & $0.010\pm 0.002$ &
1.753\\
\hline
\end{tabular}
\end{table*}
\begin{table*}
\caption{Parameter values for charged kaons(averaged) produced in
$AuAu$ collision at $\sqrt{s_{\rm{NN}}}=130$ GeV} \centering
\begin{tabular}{lllll}
\hline Centrality & $C_3$ & $\alpha$(c/GeV) & $\beta$(c/GeV)$^2$ &
$\chi^2$/ndf\\
\hline $0-5\%$ & $175\pm38$ & $0.049\pm0.009$ & $0.018\pm 0.007$
& 0.276\\
$5-15\%$ & $135\pm6$ & $0.052\pm0.009$ & $0.017\pm 0.007$ &
0.489\\
$15-30\%$ & $79\pm3$ & $0.047\pm0.008$ & $0.016\pm 0.008$ &
0.440\\
$30-60\%$ & $32\pm4$ & $0.029\pm0.006$ & $0.014\pm 0.005$ &
0.150\\
$60-90\%$ & $4.2\pm0.5$ & $0.019\pm0.008$ & $0.012\pm 0.001$ &
0.322\\
\hline
\end{tabular}
\end{table*}
\begin{table*}
\caption{Parameter values for charged kaons(averaged) produced in
$AuAu$ collision at $\sqrt{s_{\rm{NN}}}=200$ GeV} \centering
\begin{tabular}{lllll}
\hline Centrality & $C_3$ & $\alpha$(c/GeV) & $\beta$(c/GeV)$^2$ &
$\chi^2$/ndf\\
\hline $0-5\%$ & $150\pm10$ & $0.081\pm0.010$ & $0.034\pm 0.006$
& 0.622\\
$10-15\%$ & $115\pm7$ & $0.080\pm0.011$ & $0.033\pm 0.006$ &
1.163\\
$20-30\%$ & $70\pm9$ & $0.076\pm0.009$ & $0.031\pm 0.005$ &
1.067\\
$30-40\%$ & $45\pm8$ & $0.070\pm0.009$ & $0.026\pm 0.006$ &
0.659\\
$40-50\%$ & $30\pm7$ & $0.060\pm0.008$ & $0.026\pm 0.005$ &
0.806\\
$50-60\%$ & $15\pm5$ & $0.054\pm0.007$ & $0.024\pm 0.005$ &
1.441\\
$60-70\%$ & $7\pm1$ & $0.043\pm0.007$ & $0.022\pm 0.004$ &
0.996\\
$70-80\%$ & $3.0\pm0.4$ & $0.034\pm0.005$ & $0.021\pm 0.004$ &
1.762\\
$80-91\%$ & $1.4\pm0.1$ & $0.025\pm0.004$ & $0.020\pm 0.003$ &
2.015\\
\hline
\end{tabular}
\end{table*}
\begin{table*}
\caption{Parameter values for production of secondary protons in
$AuAu$ collision at $\sqrt{s_{\rm{NN}}}=130$ GeV} \centering
\begin{tabular}{lllll}
\hline Centrality & $C_3$ & $\alpha$(c/GeV) & $\beta$(c/GeV)$^2$ &
$\chi^2$/ndf\\
\hline $0-5\%$ & $23\pm5$ & $0.21\pm0.03$ & $0.045\pm 0.011$
& 1.028\\
$5-15\%$ & $16\pm3$ & $0.22\pm0.02$ & $0.053\pm 0.008$ &
0.895\\
$15-30\%$ & $12\pm2$ & $0.19\pm0.02$ & $0.042\pm 0.008$ &
0.936\\
$30-60\%$ & $6.0\pm0.3$ & $0.15\pm0.01$ & $0.031\pm 0.003$ &
0.953\\
$60-90\%$ & $1.0\pm0.1$ & $0.011\pm0.02$ & $0.023\pm 0.006$ &
2.023\\
\hline
\end{tabular}
\end{table*}
\begin{table*}
\caption{Parameter values for production of secondary protons in
$AuAu$ collision at $\sqrt{s_{\rm{NN}}}=200$ GeV} \centering
\begin{tabular}{lllll}
\hline Centrality & $C_3$ & $\alpha$(c/GeV) & $\beta$(c/GeV)$^2$ &
$\chi^2$/ndf\\
\hline $0-5\%$ & $19\pm3$ & $0.19\pm0.02$ & $0.037\pm 0.004$
& 1.526\\
$10-15\%$ & $17\pm4$ & $0.18\pm0.02$ & $0.034\pm 0.005$ &
0.927\\
$20-30\%$ & $12\pm2$ & $0.16\pm0.02$ & $0.030\pm 0.004$ &
1.085\\
$30-40\%$ & $8\pm1$ & $0.15\pm0.01$ & $0.027\pm 0.003$ &
1.001\\
$40-50\%$ & $6.5\pm0.7$ & $0.13\pm0.02$ & $0.025\pm 0.005$ &
1.703\\
$50-60\%$ & $4.0\pm0.5$ & $0.12\pm0.01$ & $0.023\pm 0.003$ &
1.494\\
$60-70\%$ & $1.9\pm0.3$ & $0.11\pm0.01$ & $0.020\pm 0.004$ &
1.896\\
$70-80\%$ & $0.8\pm0.1$ & $0.10\pm0.01$ & $0.017\pm 0.003$ &
2.116\\
$80-91\%$ & $0.50\pm0.05$ & $0.08\pm0.01$ & $0.015\pm 0.002$ &
1.839\\
\hline
\end{tabular}
\end{table*}
\begin{table*}
\caption{Parameter values for production of secondary antiprotons
in $AuAu$ collision at $\sqrt{s_{\rm{NN}}}=$130 GeV} \centering
\begin{tabular}{lllll}
\hline Centrality & $C_3$ & $\alpha$(c/GeV) & $\beta$(c/GeV)$^2$ &
$\chi^2$/ndf\\
\hline $0-5\%$ & $13\pm2$ & $0.27\pm0.02$ & $0.050\pm 0.008$
& 1.336\\
$5-15\%$ & $8\pm2$ & $0.28\pm0.03$ & $0.052\pm 0.010$ &
1.652\\
$15-30\%$ & $7.3\pm0.5$ & $0.24\pm0.01$ & $0.049\pm 0.005$ &
1.215\\
$30-60\%$ & $3.7\pm0.5$ & $0.22\pm0.01$ & $0.046\pm 0.007$ &
1.658\\
$60-90\%$ & $0.5\pm0.1$ & $0.18\pm0.02$ & $0.028\pm 0.011$ &
1.958\\
\hline
\end{tabular}
\end{table*}
\begin{table*}
\caption{Parameter values for production of secondary antiprotons
in $AuAu$ collision at $\sqrt{s_{\rm{NN}}}=200$ GeV} \centering
\begin{tabular}{lllll}
\hline Centrality & $C_3$ & $\alpha$(c/GeV) & $\beta$(c/GeV)$^2$ &
$\chi^2$/ndf\\
\hline $0-5\%$ & $8.6\pm0.7$ & $0.27\pm0.02$ & $0.048\pm 0.003$
& 1.173\\
$10-15\%$ & $8.5\pm0.8$ & $0.26\pm0.01$ & $0.046\pm 0.002$ &
1.102\\
$20-30\%$ & $5.8\pm0.7$ & $0.25\pm0.01$ & $0.043\pm 0.003$ &
1.436\\
$30-40\%$ & $4.1\pm0.6$ & $0.23\pm0.01$ & $0.042\pm 0.004$ &
1.269\\
$40-50\%$ & $3.0\pm0.3$ & $0.22\pm0.01$ & $0.040\pm 0.003$ &
1.126\\
$50-60\%$ & $2.2\pm0.3$ & $0.20\pm0.01$ & $0.039\pm 0.004$ &
1.059\\
$60-70\%$ & $1.5\pm0.1$ & $0.17\pm0.02$ & $0.034\pm 0.003$ &
1.481\\
$70-80\%$ & $0.6\pm0.1$ & $0.15\pm0.01$ & $0.033\pm 0.005$ &
1.470\\
$80-91\%$ & $0.32\pm0.05$ & $0.14\pm0.01$ & $0.029\pm 0.006$ &
1.994\\
\hline
\end{tabular}
\end{table*}
\begin{table*}
\caption{Values of different parameters to obtain the `Fit' for
$\alpha$ on the basis of eqn.(9).} \centering
\begin{tabular}{llll}
\hline Secondary type & Collision energy & $R$(c/GeV) & $S$(c/GeV)\\
\hline $\frac{h^++h^-}{2}$ & 130 GeV & 0.036 & 0.0085\\
 & 200 GeV & 0.018 & 0.0085\\
\hline $\frac{\pi^++\pi^-}{2}$ & 130 GeV & 0.023 & 0.0085\\
 & 200 GeV & 0.0003 & 0.0087\\
\hline $\frac{K^++K^-}{2}$ & 130 GeV & -0.011 & 0.011\\
 & 200 GeV & 0.001 & 0.015\\
\hline $P$ & 130 GeV & 0.052 & 0.026\\
 & 200 GeV & 0.034 & 0.027\\
\hline  & 130 GeV & 0.11 & 0.025\\
 $\bar{P}$ & 200 GeV & 0.010 & 0.026\\
\hline
\end{tabular}
\end{table*}
\begin{table*}
\caption{Values of different parameters to obtain the `Fit' for
$\beta$ on the basis of eqn.(9).} \centering
\begin{tabular}{llll}
\hline Secondary type & Collision energy & $R$(c/GeV)$^2$ & $S$(c/GeV)$^2$\\
\hline $\frac{h^++h^-}{2}$ & 130 GeV & 0.0035 & 0.0021\\
 & 200 GeV & 0.0001 & 0.0019\\
\hline $\frac{\pi^++\pi^-}{2}$ & 130 GeV & 0.008 & 0.0019\\
 & 200 GeV & 0.0055 & 0.0019\\
\hline $\frac{K^++K^-}{2}$ & 130 GeV & 0.006 & 0.0019\\
 & 200 GeV & 0.016 & 0.0020\\
\hline $P$ & 130 GeV & 0.012 & 0.005\\
 & 200 GeV & 0.0039 & 0.0048\\
\hline  & 130 GeV & 0.022 & 0.0051\\
 $\bar{P}$ & 200 GeV & 0.020 & 0.0048\\
\hline
\end{tabular}
\end{table*}

\newpage

\begin{figure*}
\subfigure[]{
\begin{minipage}{.5\textwidth}
\centering
\includegraphics[width=11cm]{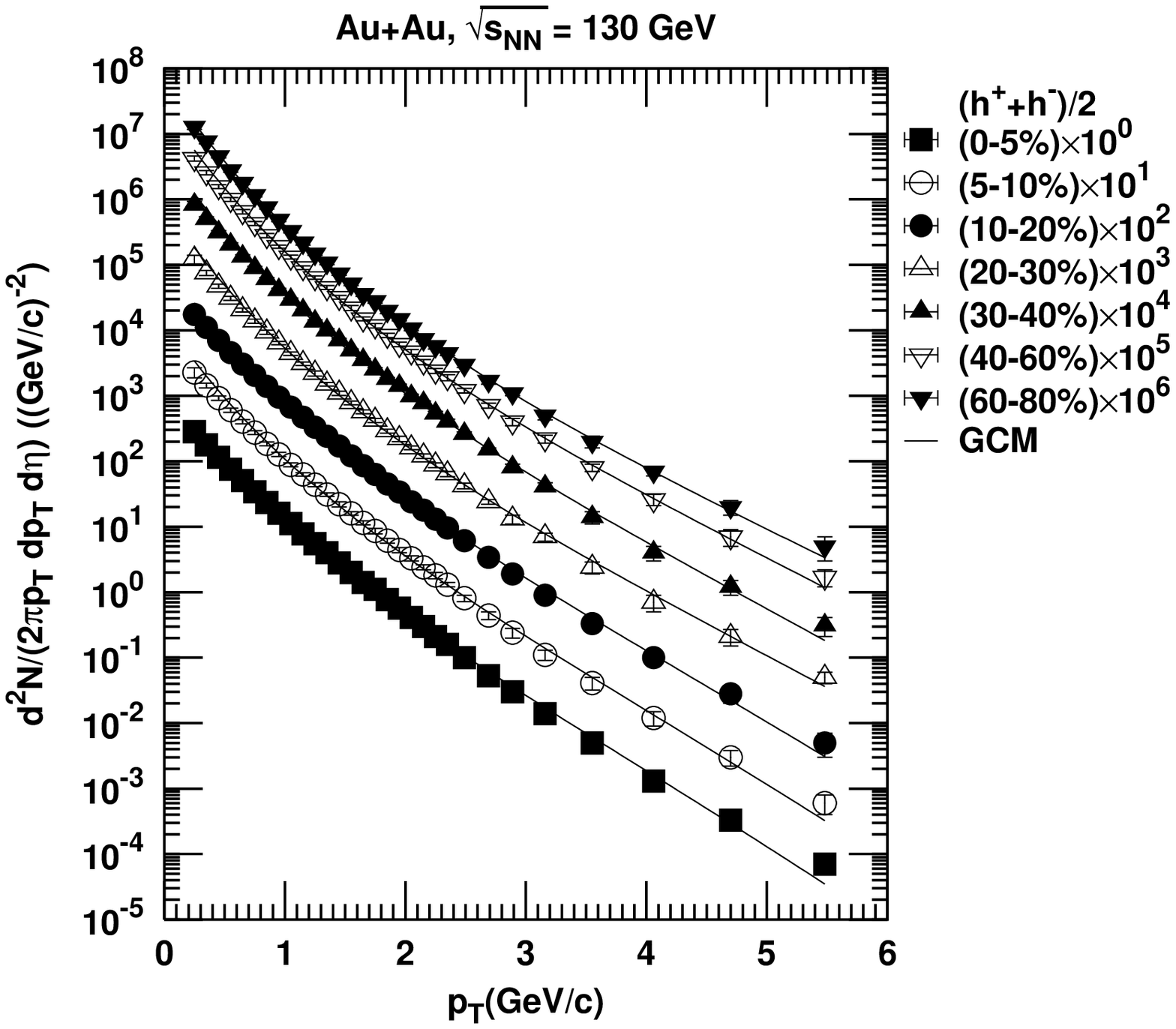}
\end{minipage}}%
\subfigure[]{
\begin{minipage}{.5\textwidth}
\centering
\includegraphics[width=11cm]{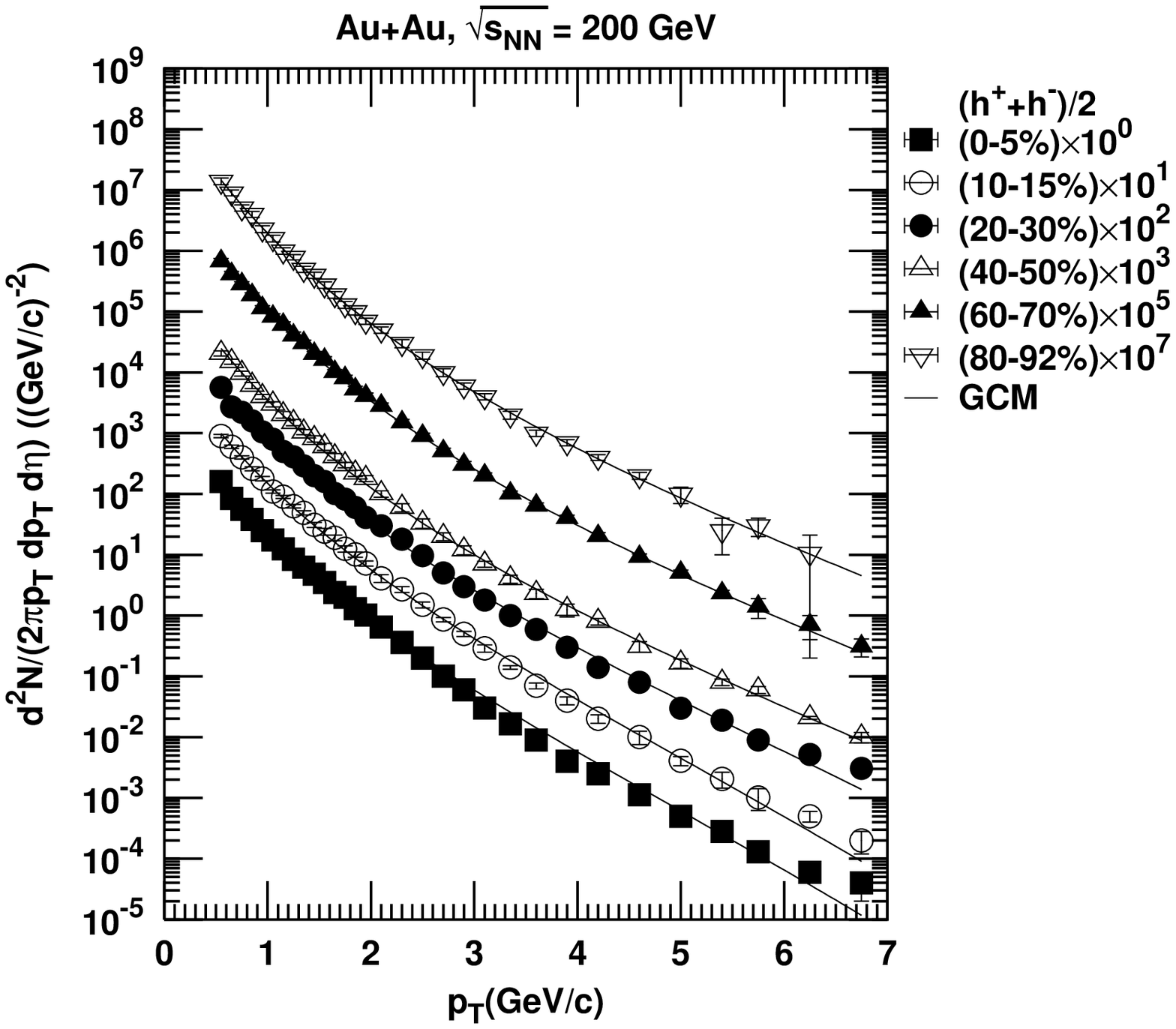}
\end{minipage}}%
\caption{Plots of invariant spectra of secondary charged hadrons
produced in $AuAu$ collisions at two different RHIC-energies for
various centrality-bins. The experimental data points are taken
from Ref.\cite{Adler1} for $\sqrt{s_{\rm{NN}}}=130$ GeV and from
Ref.\cite{Jia1} for $\sqrt{s_{\rm{NN}}}=200$ GeV. The solid curves
provide the GCM-based fits.}
\subfigure[]{
\begin{minipage}{.5\textwidth}
\centering
\includegraphics[width=11cm]{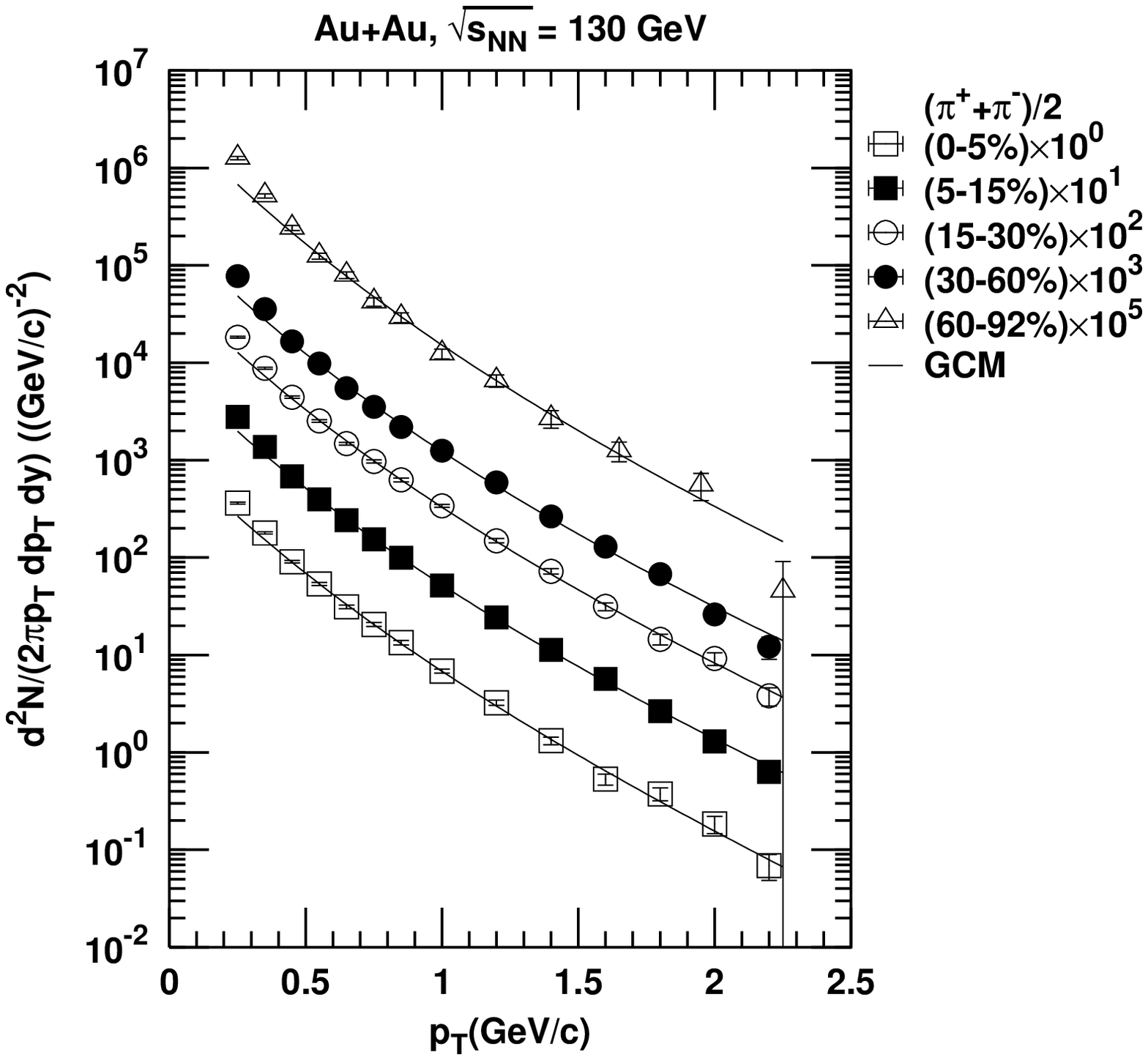}
\end{minipage}}%
\subfigure[]{
\begin{minipage}{.5\textwidth}
\centering
\includegraphics[width=11cm]{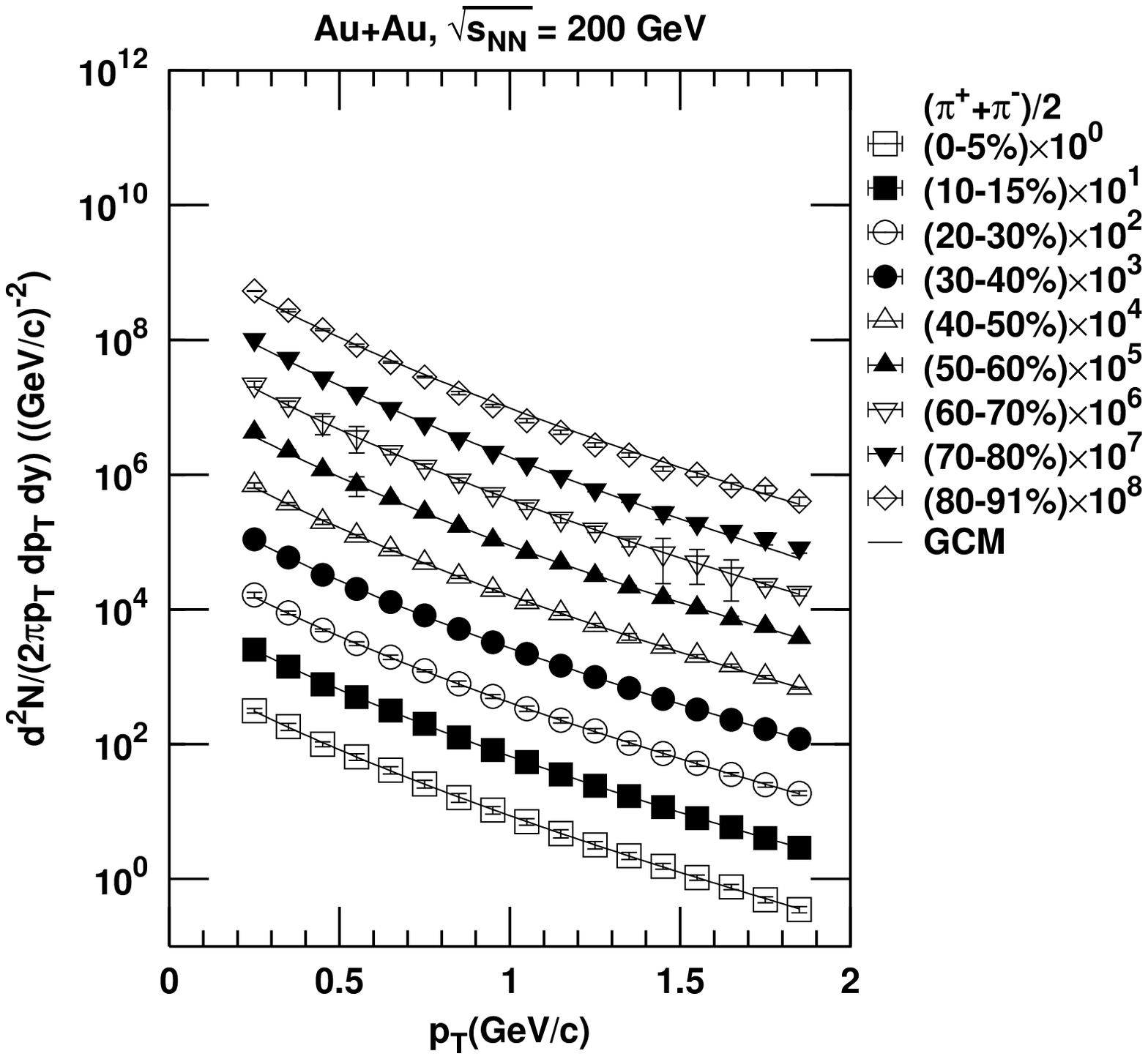}
\end{minipage}}%
\caption{Nature of invariant spectra of secondary charged pions
produced in $AuAu$ collisions at two different RHIC-energies at
different centralities as a function of $p_{\rm{T}}$. The experimental
data points at $\sqrt{s_{\rm{NN}}}=130$ GeV are taken from
Ref.\cite{Hoy1} while those for $\sqrt{s_{\rm{NN}}}=200$ GeV are from
Ref.\cite{Chujo1}. The solid curves are drawn on the basis of
eqn.(6).}
\end{figure*}
\begin{figure*}
\subfigure[]{
\begin{minipage}{.5\textwidth}
\centering
\includegraphics[width=11cm]{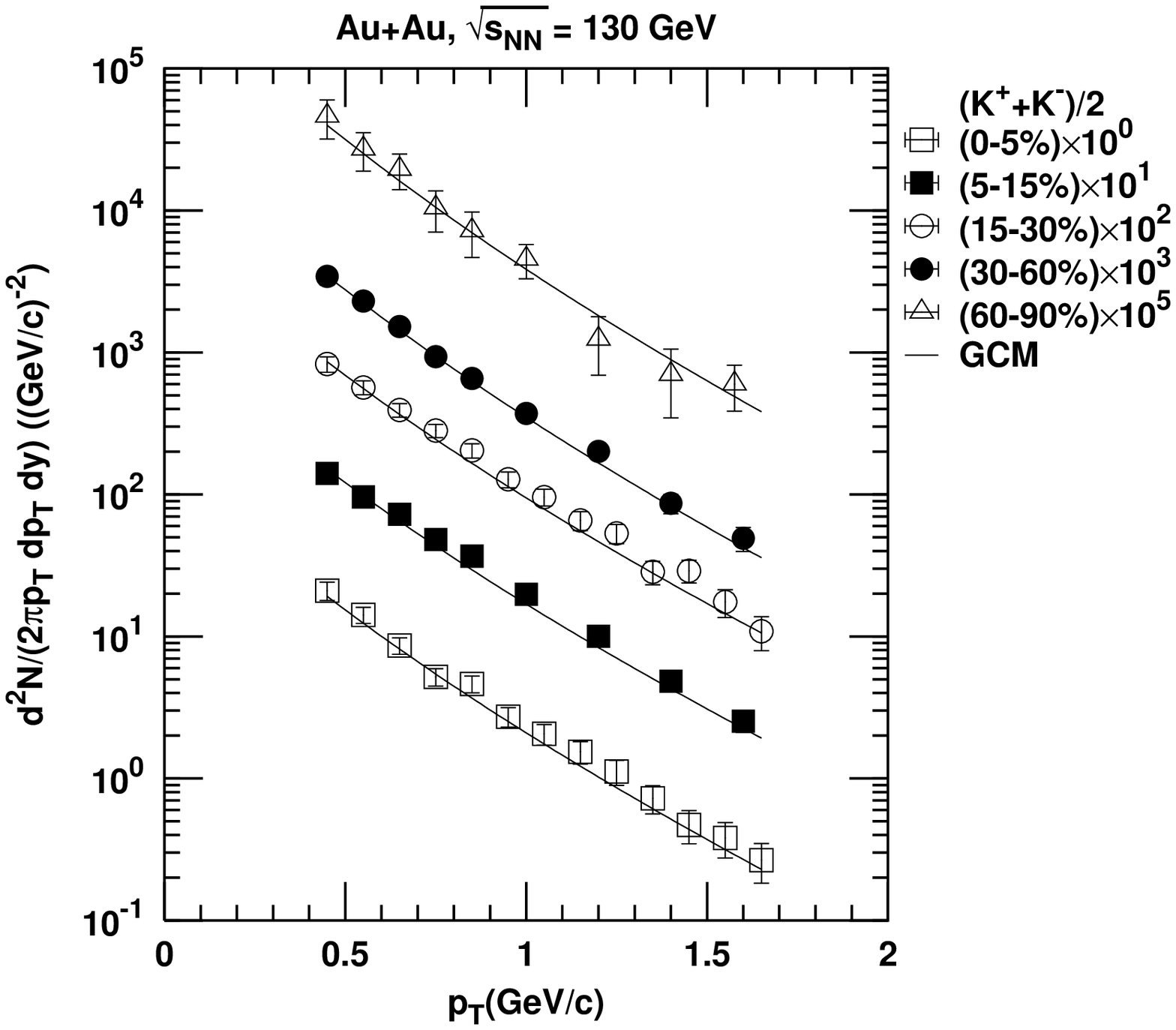}
\end{minipage}}%
\subfigure[]{
\begin{minipage}{.5\textwidth}
\centering
\includegraphics[width=11cm]{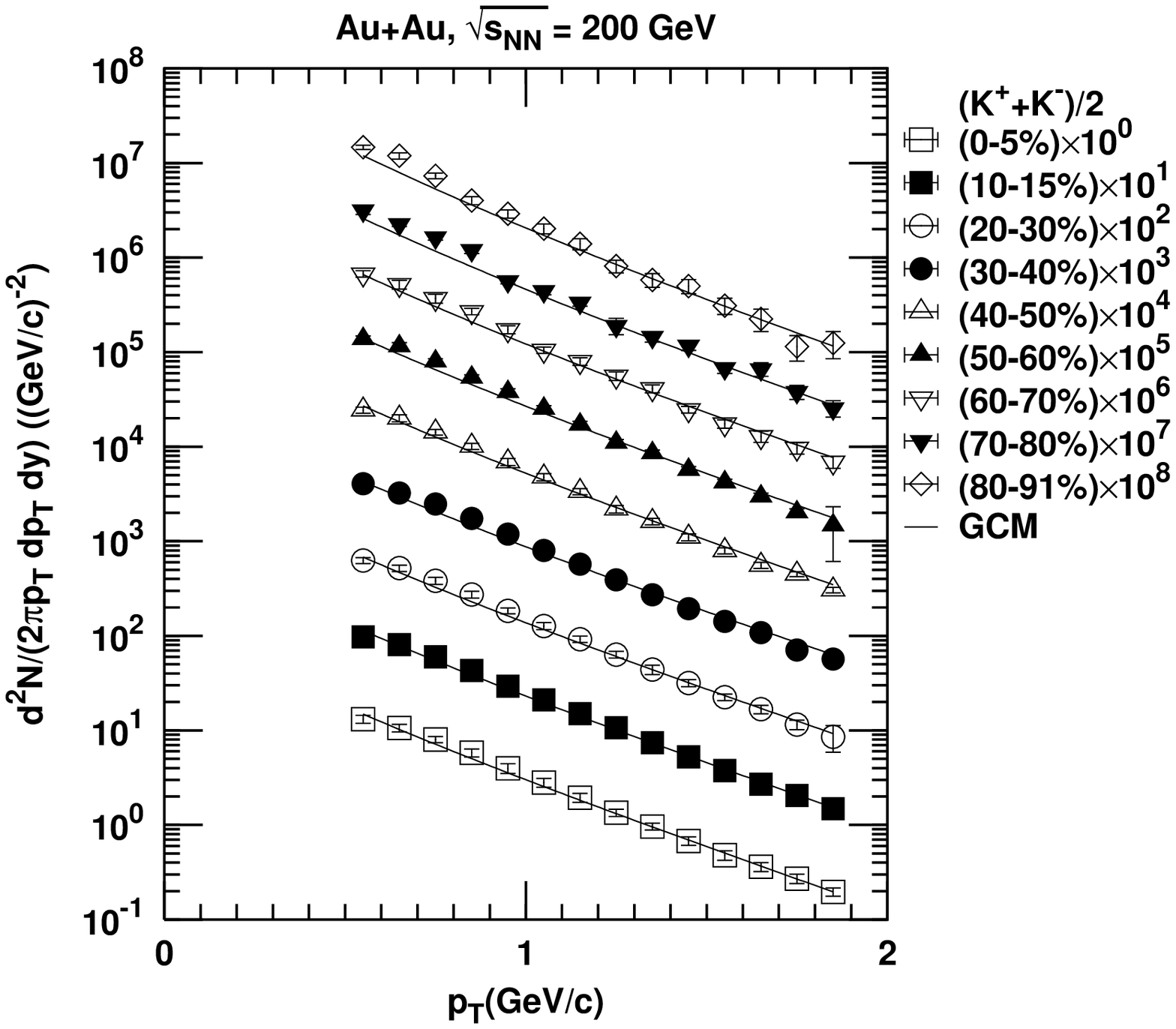}
\end{minipage}}%
\caption{Transverse momentum spectra for production of secondary
charged kaons in $AuAu$ collisions at two different RHIC-energies
at different centralities. The various experimental data points at
$\sqrt{s_{\rm{NN}}}=130$ GeV are taken from Ref.\cite{Hoy1} while those
for $\sqrt{s_{\rm{NN}}}=200$ GeV are from Ref.\cite{Chujo1}. The
present model(GCM)-based fits are depicted by the solid
curvilinear lines.}
\subfigure[]{
\begin{minipage}{.5\textwidth}
\centering
\includegraphics[width=11cm]{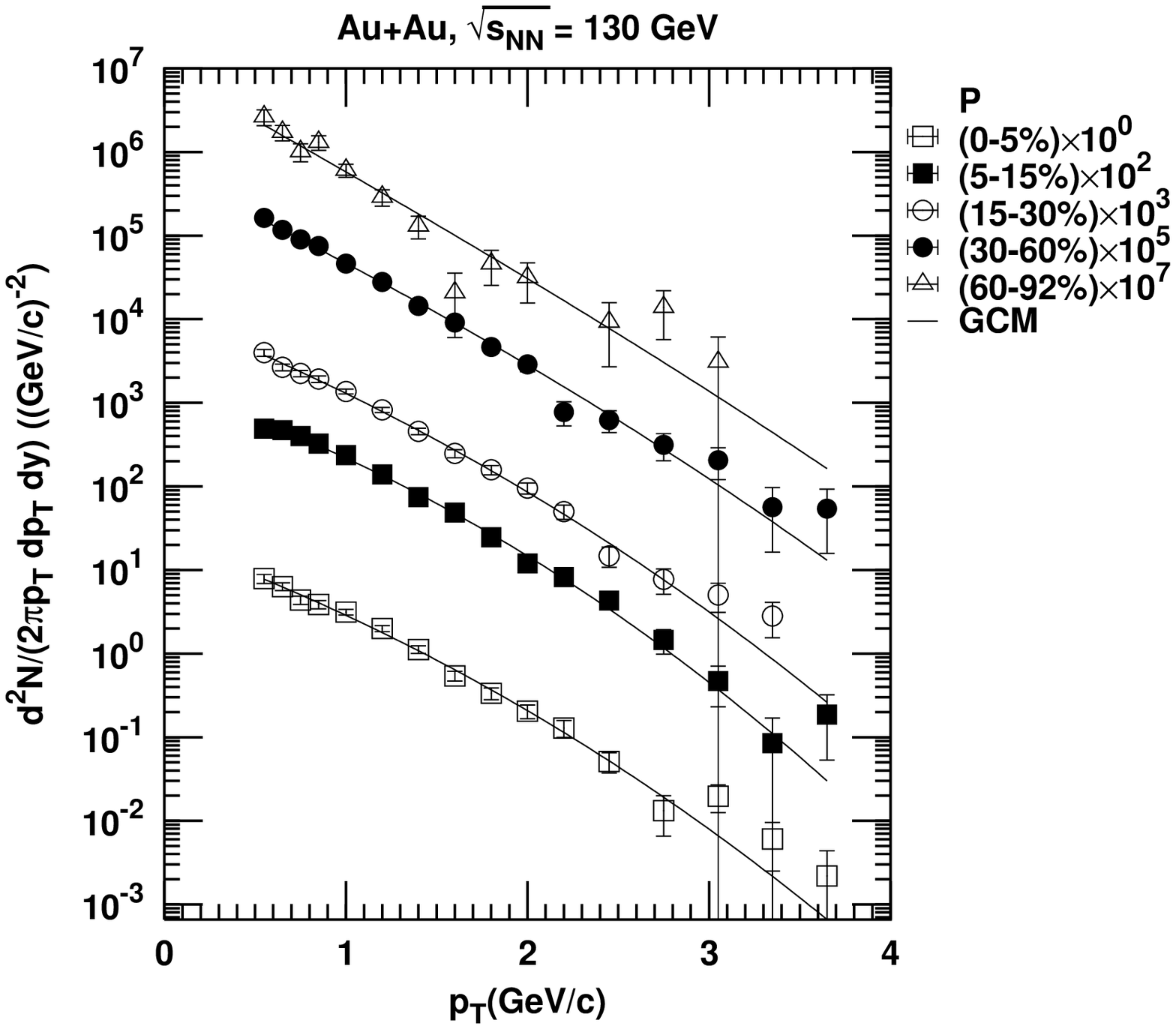}
\end{minipage}}%
\subfigure[]{
\begin{minipage}{.5\textwidth}
\centering
\includegraphics[width=11cm]{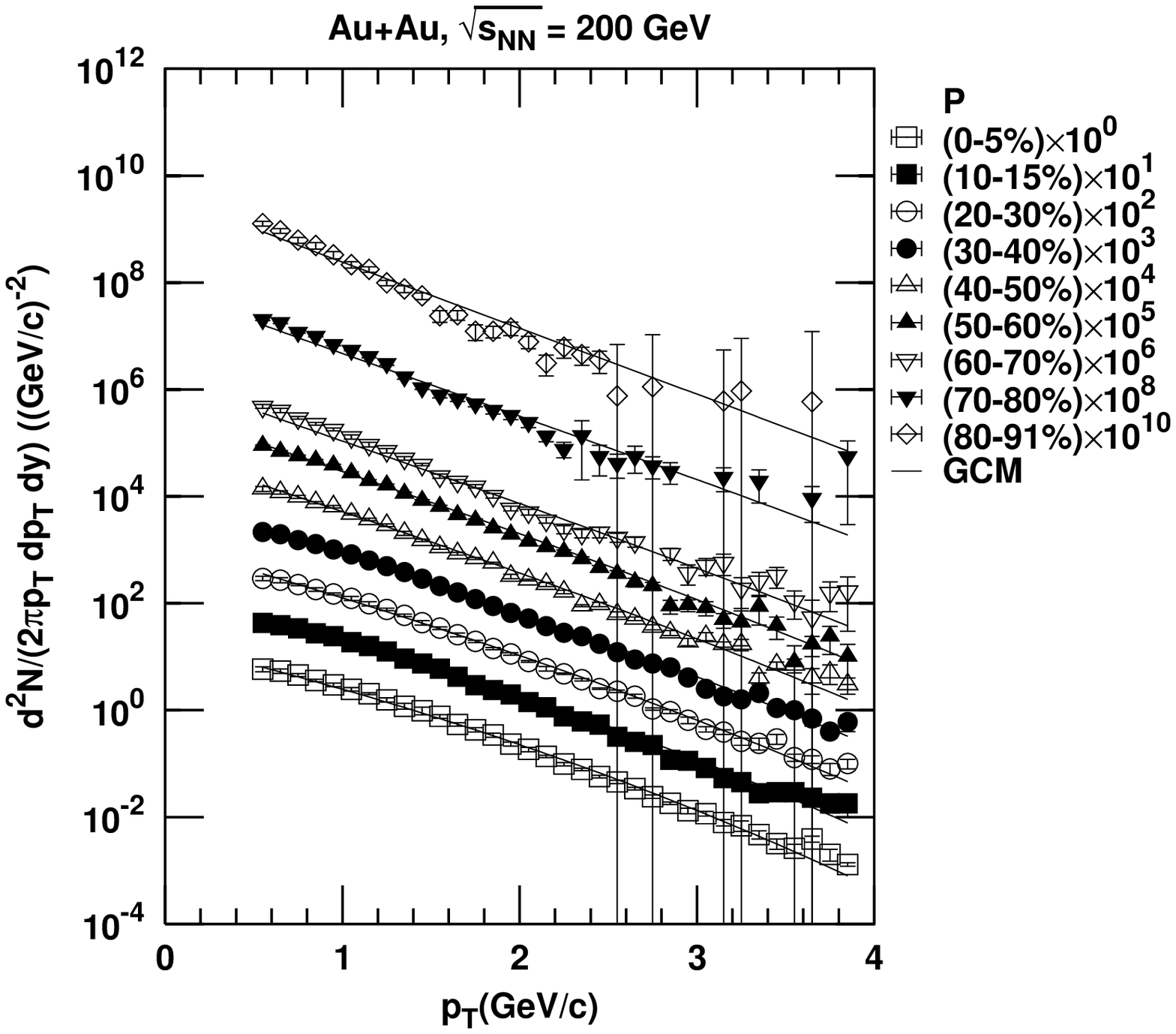}
\end{minipage}}%
\caption{Plots of invariant spectra for production of secondary
protons in $AuAu$ collisions at two different RHIC-energies at
different centralities. The various experimental data points at
$\sqrt{s_{\rm{NN}}}=130$ GeV are taken from Ref.\cite{Hoy1} while those
for $\sqrt{s_{\rm{NN}}}=200$ GeV are from Ref.\cite{Chujo1}. The
GCM-based fits are depicted by the solid curves.}
\end{figure*}
\begin{figure*}
\subfigure[]{
\begin{minipage}{0.5\textwidth}
\centering
\includegraphics[width=11cm]{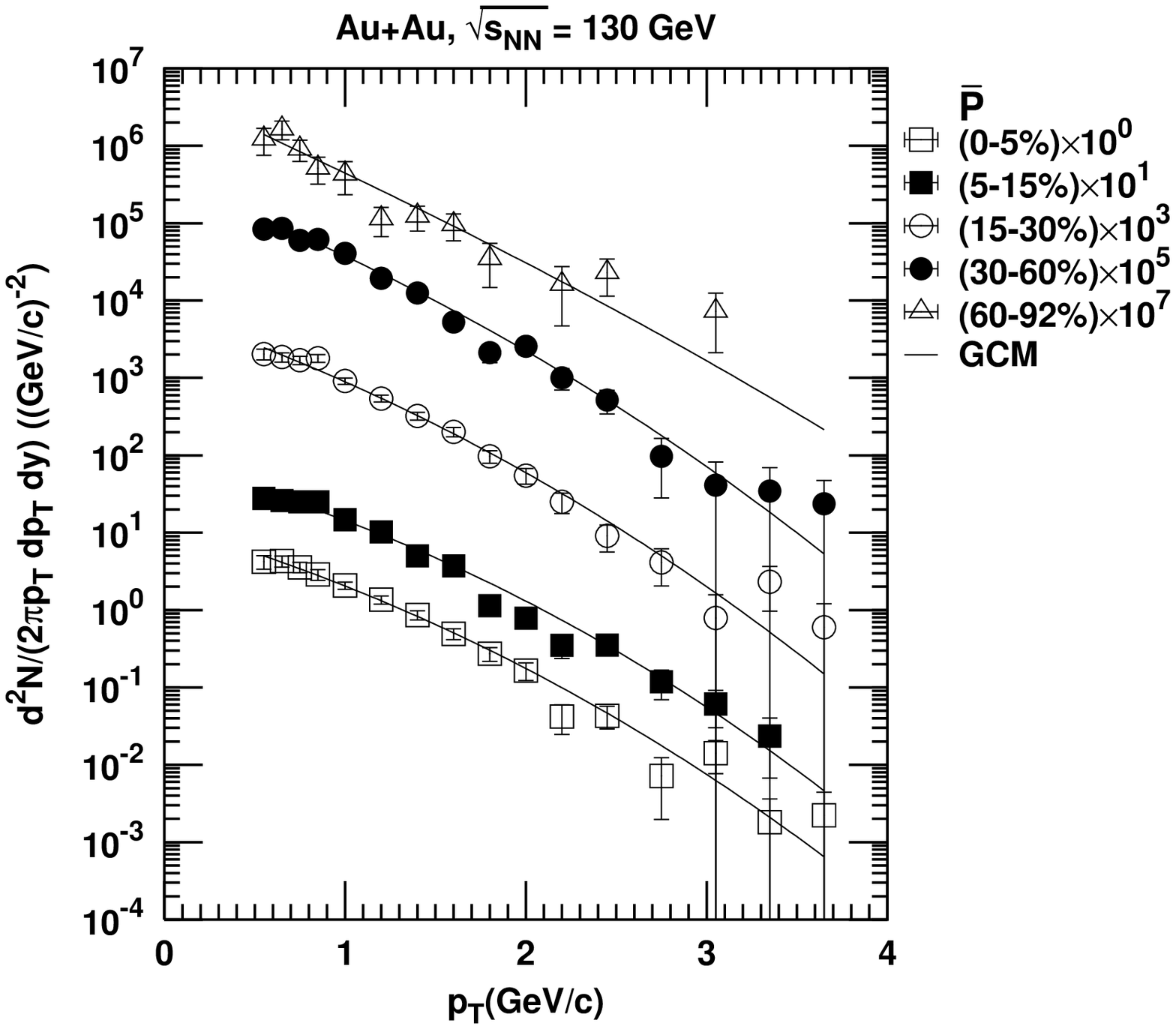}
\end{minipage}}%
\subfigure[]{
\begin{minipage}{0.5\textwidth}
\centering
\includegraphics[width=11cm]{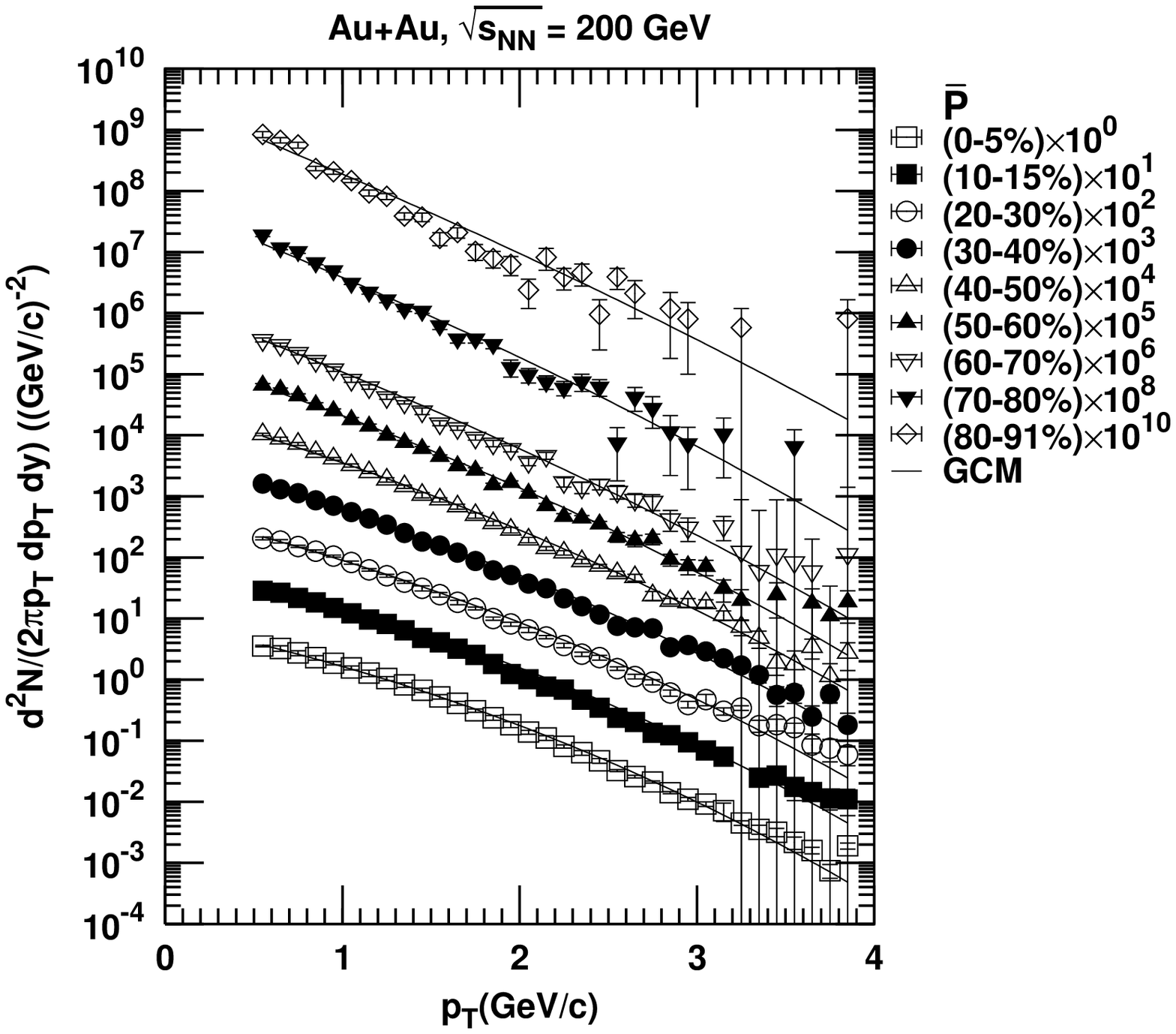}
\end{minipage}}%
\caption{Nature of invariant spectra of secondary antiprotons
produced in $AuAu$ collisions at two different RHIC-energies at
different centralities as a function of $p_{\rm{T}}$. The experimental
data points at $\sqrt{s_{\rm{NN}}}=130$ GeV are taken from
Ref.\cite{Hoy1} while those for $\sqrt{s_{\rm{NN}}}=200$ GeV are from
Ref.\cite{Chujo1}. The solid curves are drawn on the basis of
eqn.(6).}
\end{figure*}
\begin{figure*}
\subfigure[]{
\begin{minipage}{.33\textwidth}
\centering
\includegraphics[width=6cm]{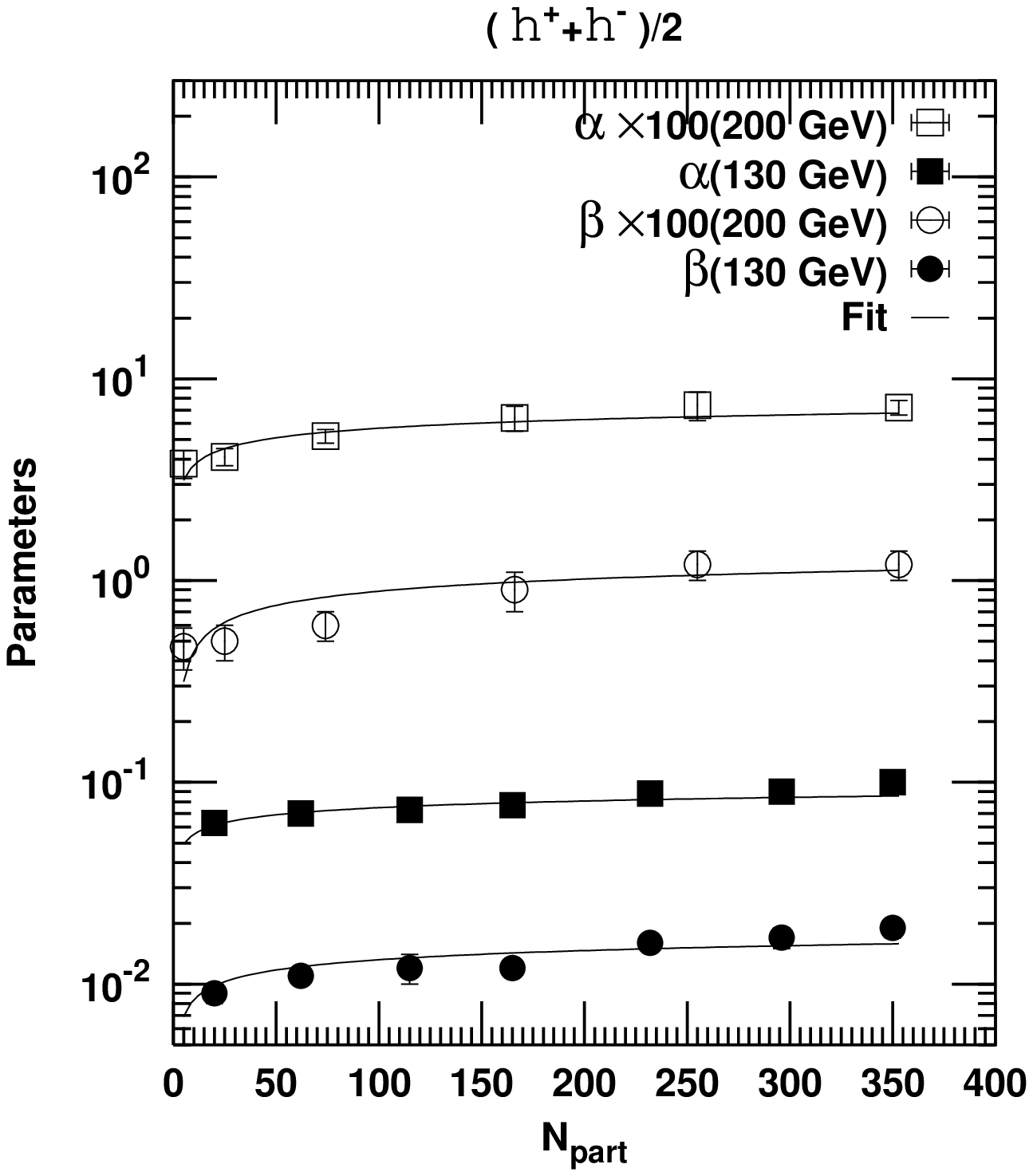}
\end{minipage}}%
\subfigure[]{
\begin{minipage}{.33\textwidth}
\centering
\includegraphics[width=6cm]{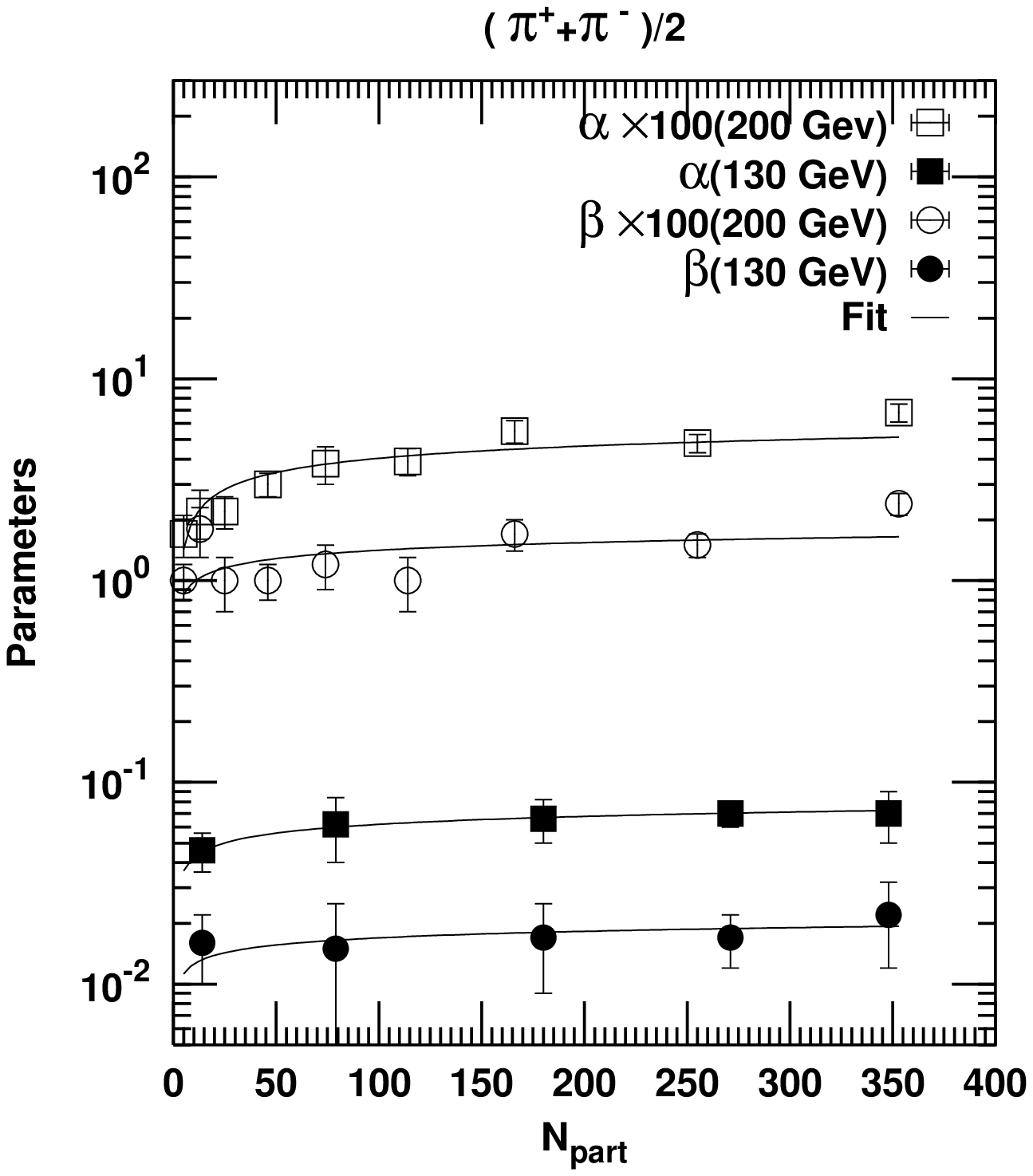}
\end{minipage}}%
\subfigure[]{
\begin{minipage}{.33\textwidth}
\centering
\includegraphics[width=6cm]{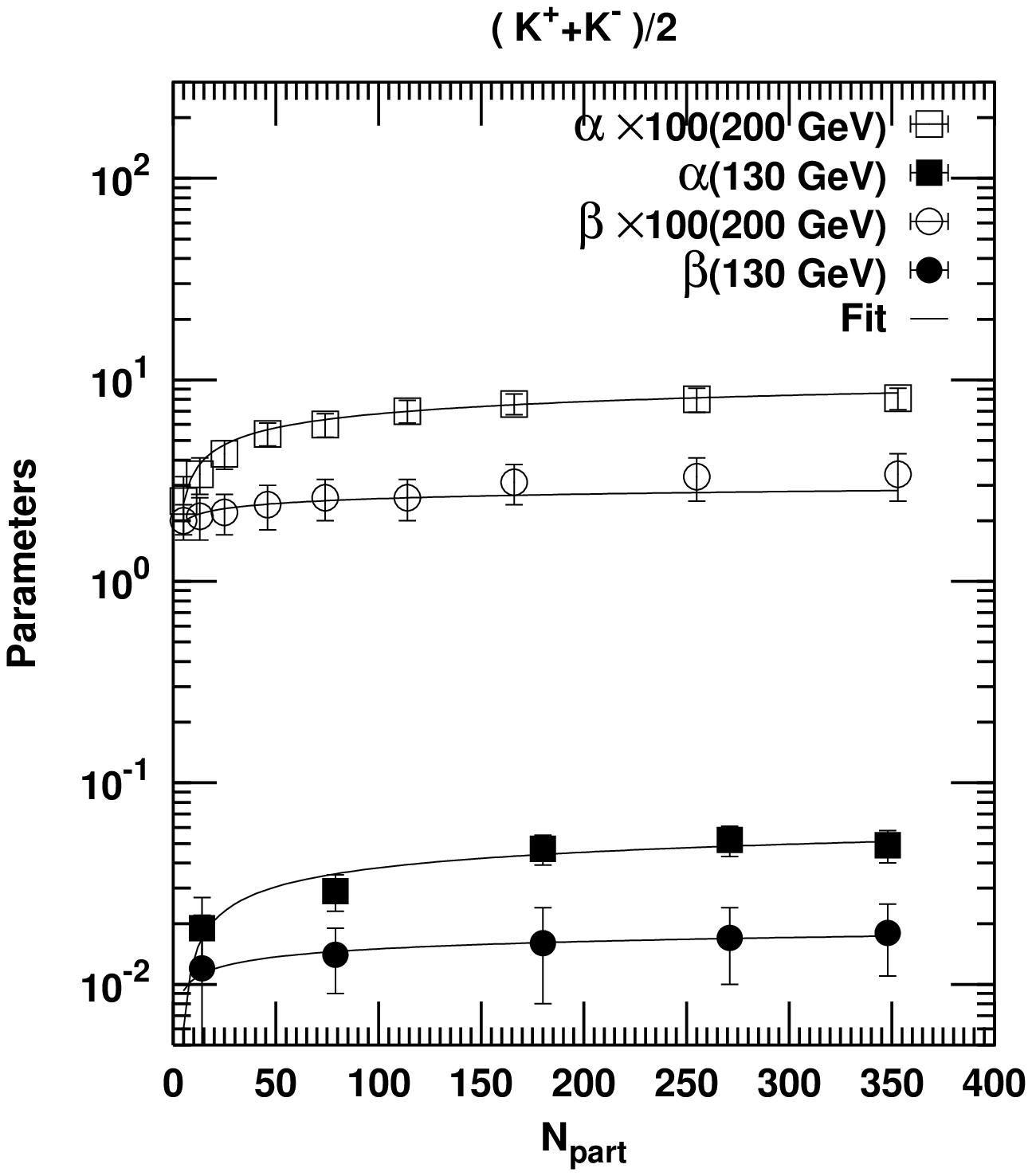}
\end{minipage}}%
\vspace{.1cm} \subfigure[]{
\begin{minipage}{.5\textwidth}
\centering
\includegraphics[width=6cm]{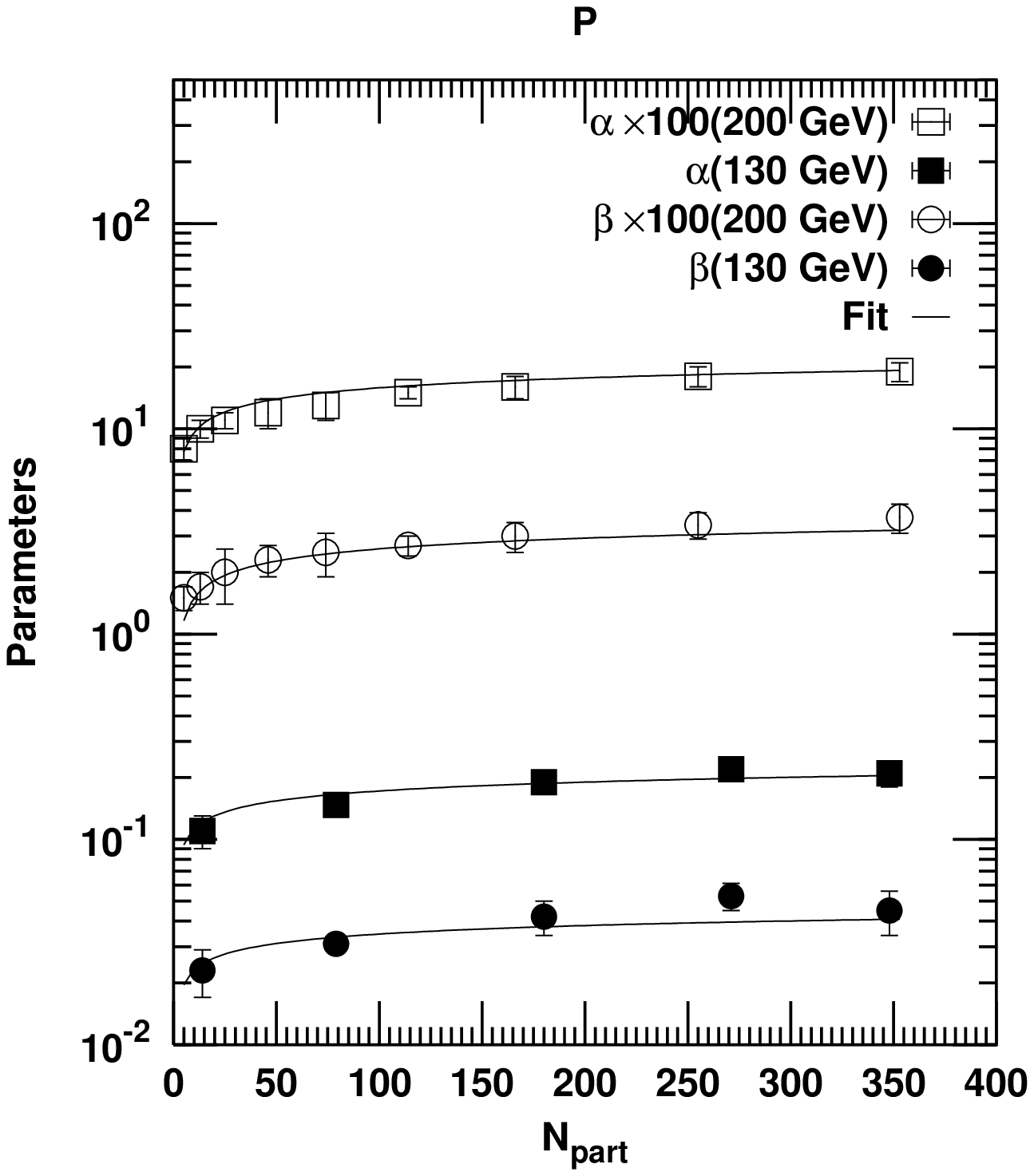}
\end{minipage}}%
\subfigure[]{
\begin{minipage}{.5\textwidth}
\centering
\includegraphics[width=6cm]{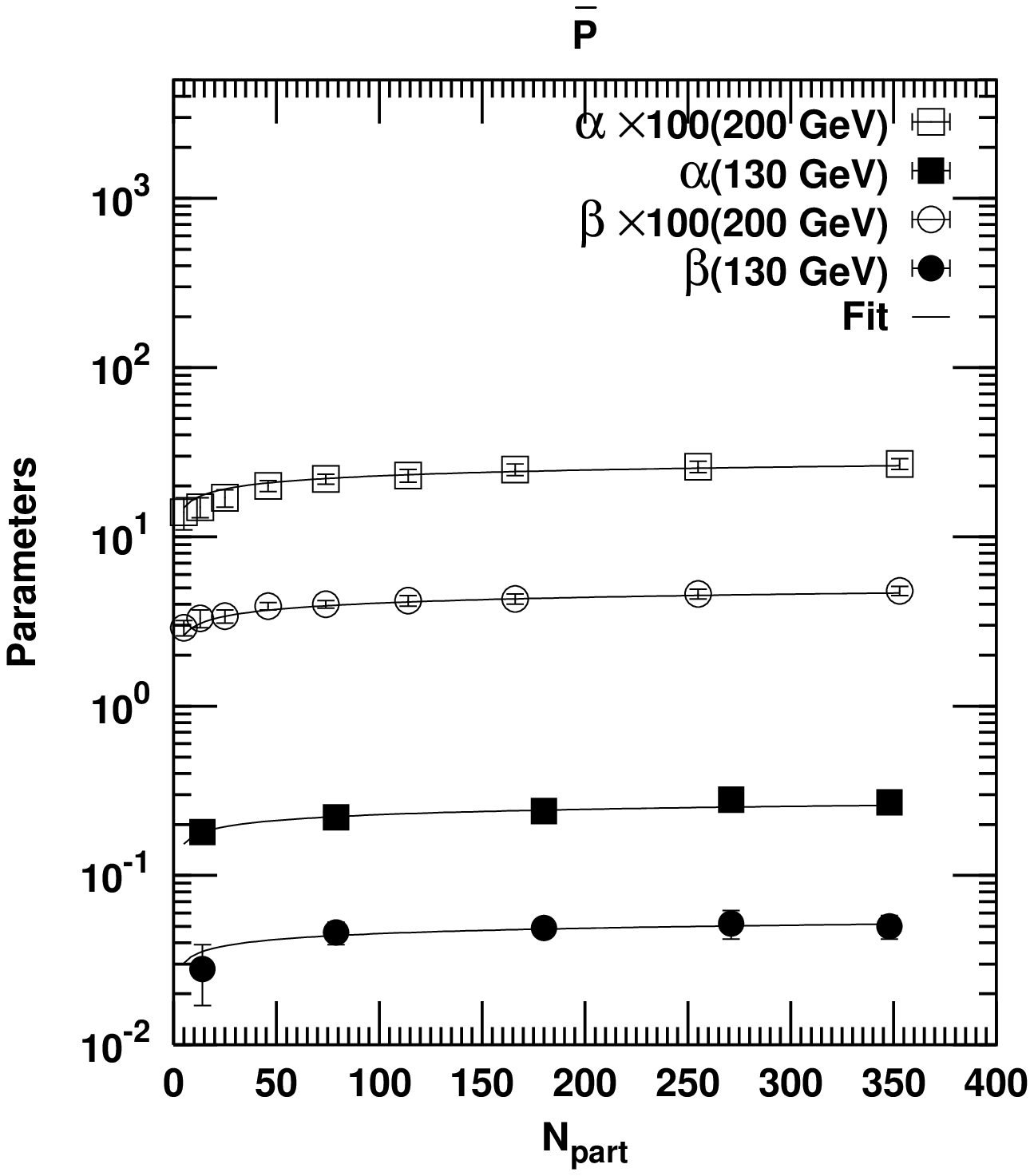}
\end{minipage}}%
\caption{Plots of $\alpha$[in (c/GeV)] and $\beta$[in (c/GeV)$^2$]
as a function of number of participant nucleons, $N_{\rm{part}}$. The
different data-type points for various secondaries are taken from
Table 2 - Table 11. Various solid curves are drawn on the basis of
eqn.(9) and denoted as `Fit' in the figures.}
\end{figure*}
\begin{figure*}
\subfigure[]{
\begin{minipage}{.5\textwidth}
\centering
\includegraphics[width=11cm]{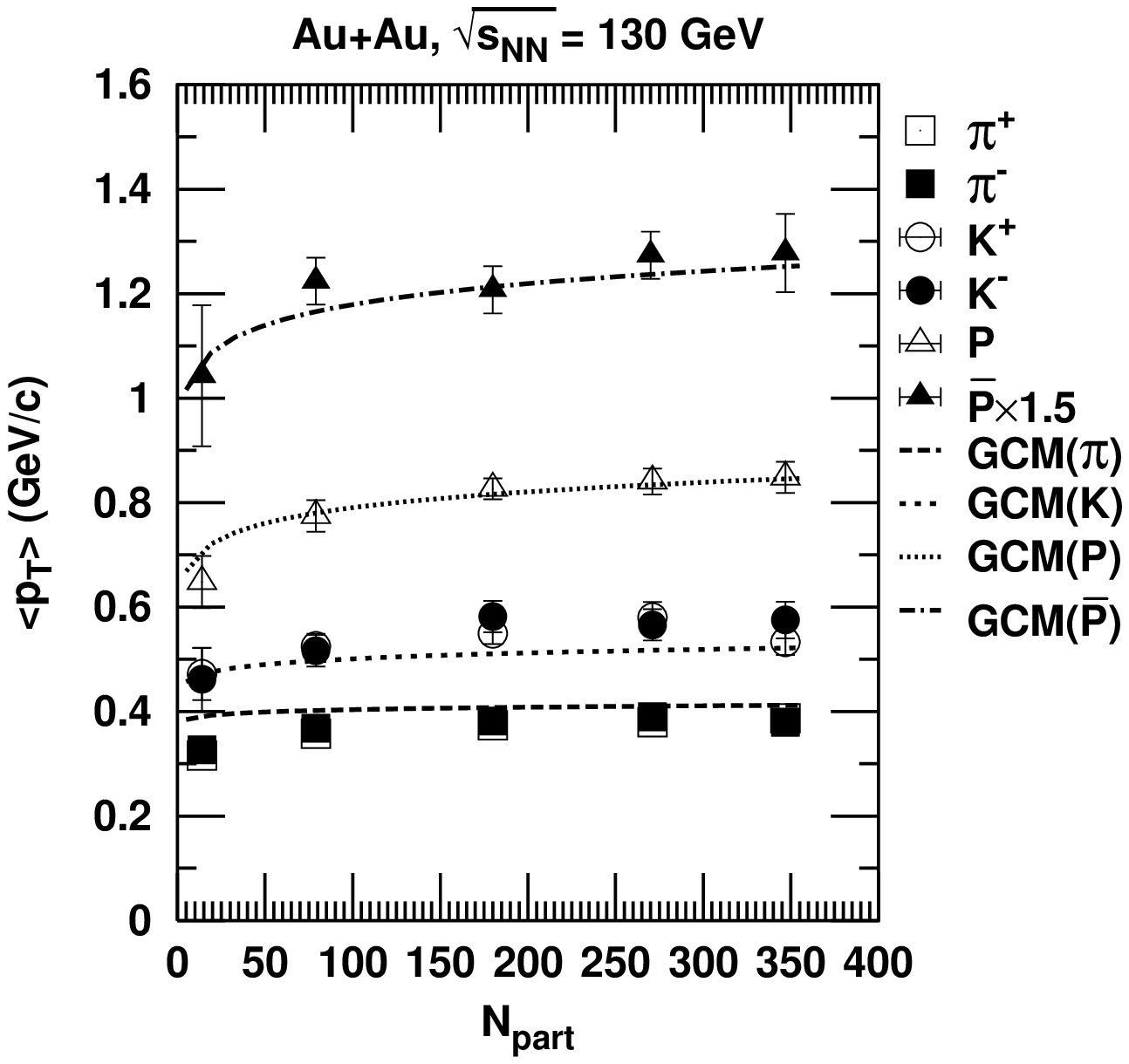}
\end{minipage}}%
\subfigure[]{
\begin{minipage}{.5\textwidth}
\centering
\includegraphics[width=11cm]{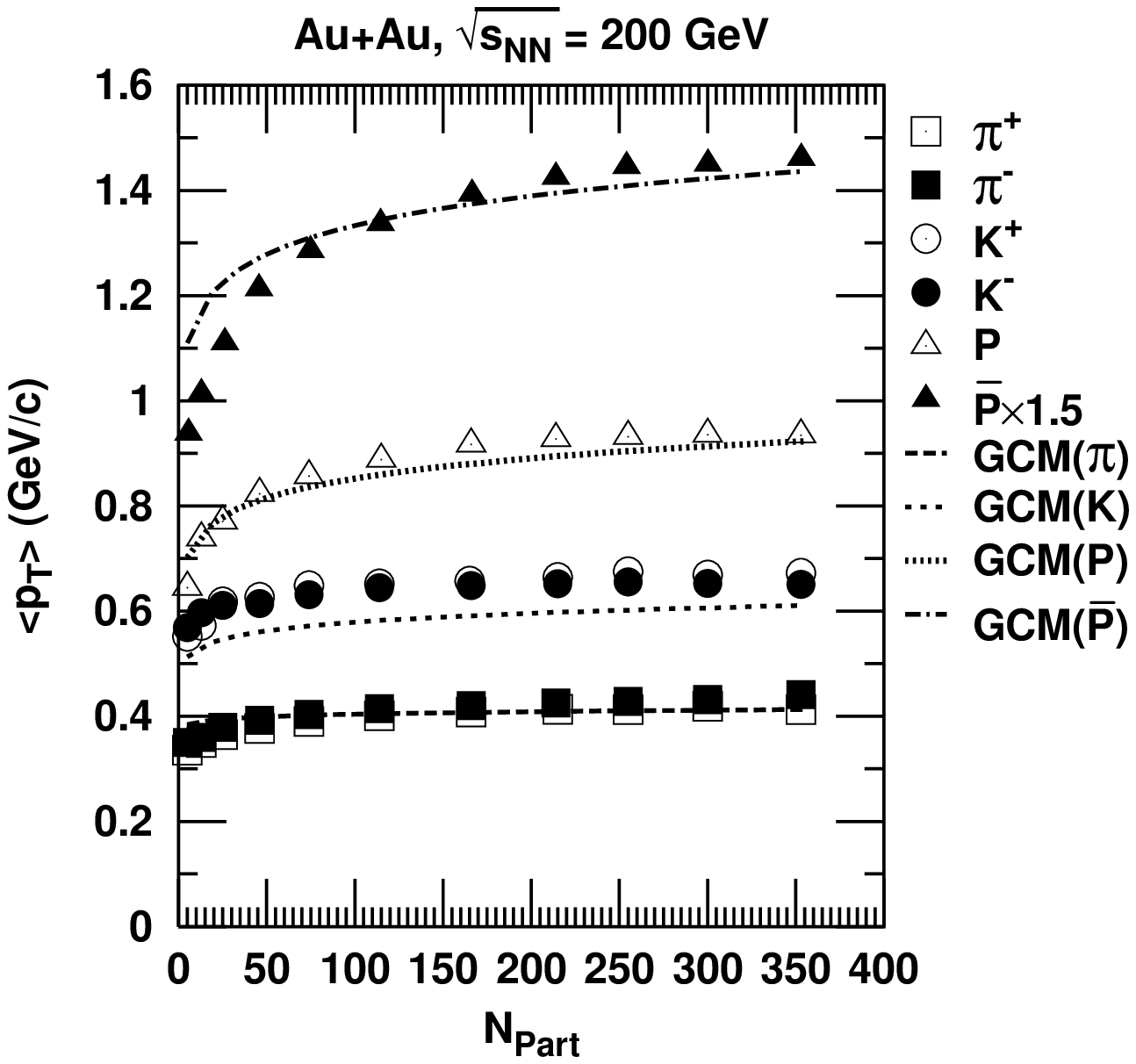}
\end{minipage}}%
\caption{Nature of average transverse momenta($<p_{\rm{T}}>$) for various
secondaries produced in $AuAu$ collisions at two different
energies as a function of $N_{\rm{part}}$. The data-type points are the
extracted-results obtained by RHIC experimental
groups\cite{Chujo1}. The GCM-based results are shown by various
dashed and dotted curves.}
\end{figure*}
\end{document}